\documentclass{egpubl}
\usepackage{eg2022}
 
\ConferencePaper        
\usepackage[T1]{fontenc}
\usepackage{dfadobe}  

\biberVersion
\BibtexOrBiblatex
\usepackage[backend=biber,bibstyle=EG,citestyle=alphabetic,backref=true]{biblatex} 
\electronicVersion
\PrintedOrElectronic

\ifpdf \usepackage[pdftex]{graphicx} \pdfcompresslevel=9
\else \usepackage[dvips]{graphicx} \fi

\usepackage{egweblnk} 

\title{Computational Design of Kinesthetic Garments}

\author[V. Vechev, J. Zarate, B. Thomaszewski, and O. Hilliges]
{\parbox{\textwidth}{\centering V.Vechev\orcid{0000-0002-1328-153X} \ 
         J. Zarate\orcid{0000-0001-9106-2394} \ 
         B. Thomaszewski\orcid{0000-0001-8086-7664} \ 
         O. Hilliges\orcid{0000-0002-5068-3474}
        }
        \\
{\parbox{\textwidth}{\centering Department of Computer Science, ETH Z{\"u}rich, Switzerland\\}
}
}

\usepackage{xcolor}

\newcommand\REV[1]{\textcolor{black}{#1}}

\newcommand\remove[1]{}

\usepackage[utf8]{inputenc}

\usepackage{tabularx}
\usepackage[percent]{overpic}

\usepackage{nameref}

\usepackage{hyperref}
\usepackage{url}

\usepackage{xspace}
\usepackage{enumitem}
\usepackage{mathtools}
\usepackage{commath}

\usepackage{amssymb}
\usepackage{pifont}

\newcommand{\customtilde}{{\raise.17ex\hbox{$\scriptstyle\sim$}}}

\let\set=\mathcal

\newcommand{\N}{\mbox{\rm \hbox{I\kern-.15em\hbox{N}}}}
\newcommand{\R}{\mbox{\rm \hbox{I\kern-.15em\hbox{R}}}}

\def \N {\mbox{\rm \hbox{I\kern-.15em\hbox{N}}}}
\def \R {\mbox{\rm \hbox{I\kern-.15em\hbox{R}}}}

\newcommand{\bC}{\mathbf{C}}

\newcommand{\bF}{\mathbf{F}}

\newcommand{\bd}{\mathbf{d}}
\newcommand{\be}{\mathbf{e}}
\newcommand{\bff}{\mathbf{f}}

\newcommand{\bn}{\mathbf{n}}

\newcommand{\bv}{\mathbf{v}}
\newcommand{\bx}{\mathbf{x}}

\newcommand{\energydensity}{W}
\newcommand{\materialdensity}{d}
\newcommand{\sensitivity}{\alpha}

\begin{document}

\teaser{
 \includegraphics[width=\linewidth]{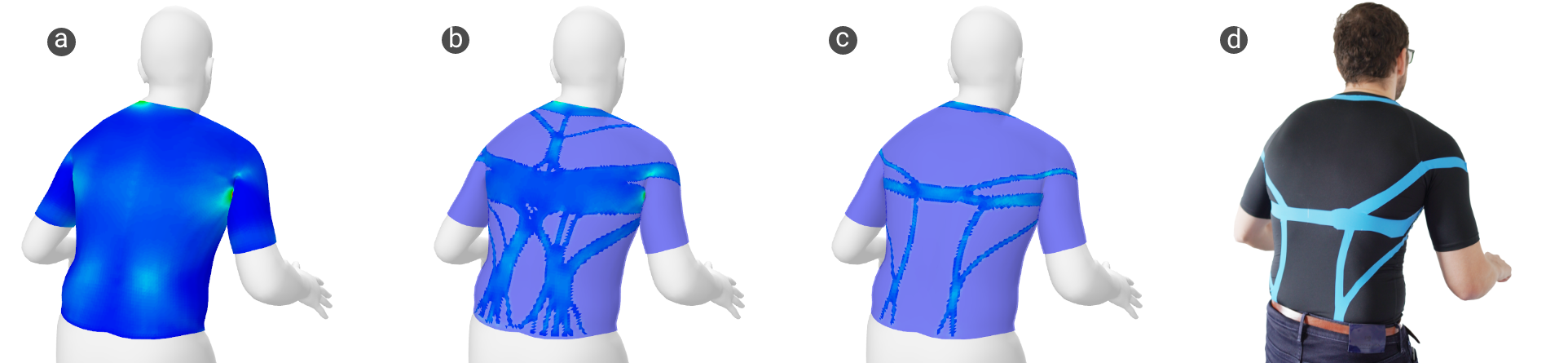}
 \centering
  \caption {Given a pose and garment, our method leverages an FEM simulation of the garment (light blue) as it deforms over the body (a) and optimizes the placement of a reinforced material (dark blue) by maximizing energy density for a given area budget. Our algorithm produces a continuous range of designs with various amounts of reinforcement coverage (b,c), allowing designers to explore the trade-off between instrumentation and performance. The selected design is then manufactured using a standard HTV (heat-transfer-vinyl) process. \REV{The above design is a posture-correcting shirt that pulls back on the user when slouching forward.}} 
\label{fig:teaser}
}


\maketitle
\begin{abstract}
Kinesthetic garments provide physical feedback on body posture and motion through tailored distributions of reinforced material. Their ability to selectively stiffen a garment's response to specific motions makes them appealing for rehabilitation, sports, robotics, and many other application fields.
However, finding designs that distribute a given amount of reinforcement material to maximally stiffen the response to specified motions is a challenging problem. In this work, we propose an optimization-driven approach for automated design of reinforcement patterns for kinesthetic garments. Our main contribution is to cast this design task as an on-body topology optimization problem. Our method allows designers to explore a continuous range of designs corresponding to various amounts of reinforcement coverage. Our model captures both tight contact and lift-off separation between cloth and body. 
We demonstrate our method on a variety of reinforcement design problems for different body sites and motions. Optimal designs lead to a two- to threefold improvement in performance in terms of energy density. A set of manufactured designs were consistently rated as providing more resistance than baselines in a comparative user study. \\


\end{abstract}  

\section{Introduction}
Kinesthetic feedback during body motion has wide-ranging applications in posture-correction, locomotion assistance \cite{kim2019reducing, lee2018autonomous}, and enhanced immersion in mixed reality \cite{gunther2019pneumact, rognon2018flyjacket}. In many recent works, \textit{wearable compliant interfaces} have emerged as the preferred means of transmitting forces to the human body, owing to their lightweight and conforming properties. An important consideration in these interfaces is to provide feedback or assistance for specific motions, while not overly instrumenting the user. However, unlike their stiff and rigid counterparts, wearable compliant interfaces behave in a non-linear and difficult to predict manner, creating a challenging problem for designers.


The goal of this work is to facilitate the design of \textit{Kinesthetic garments}---lightweight and compliant apparel that, when deformed during body motion, deliver kinesthetic feedback to the user via forces felt in the muscles. This is accomplished by reinforcing the garment with stiffer material in order to resist specific motions. In this task, designers must balance the conflicting goals of resisting specific motions and retaining as much of the garment's flexibility as possible. 
We propose to combine these objectives into the task of \textit{maximizing design efficiency}: only the minimal amount of reinforcement material required to achieve a desired stiffening effect should be used. Conversely, a given budget of material should be distributed such as to maximally resist the specified motion. 

The challenge of designing garments that consider body motion is an emerging topic in computer graphics  \cite{liu2021knitting, montes2020computational} and wearable robotics \cite{ortiz2017energy, sanchez2021textile}. A number of technical challenges arise: both the cloth and the body behave in a complex non-linear manner, and their coupling under tight contact is difficult to model. In addition, the human body itself deforms significantly under kinematic motion (i.e. muscle bulging). Optimizing for particular material budgets in an on-body and compliant setting has yet to be considered. Thus, the task of designing garments that provide kinesthetic feedback under desired motions remains difficult and time-consuming for non-expert designers.


Our idea is to cast this design task as an \textit{on-body topology optimization} problem in which we seek to compute optimal layouts for garment reinforcements. Based on this idea, we propose an automatic design algorithm that, given a material budget and specific motion will maximize the energy needed to deform the garment, and thus the mechanical work users generate during the motion and the kinesthetic feedback they experience. 
As the desired material budget may not be known a-priori, our method supports designers by generating a range of distinct designs, enabling the exploration of the trade-off between material budget and resulting performance.

As with most topology optimization methods, our approach relies on finite element analysis during optimization. For this purpose, we create a customized garment-on-body model of skin-tight cloth that is elastically stretched over a body. 
This model allows cloth to smoothly slide on the body, resulting in garments that fit tightly to convex parts of the body. It also allows for lift-off separation between cloth and body, and exhibits close-to-zero stiffness under compression to emulate wrinkling. 

We leverage our model and automatic design method to produce a number of kinesthetic garment designs for different body sites and motions. Optimal designs are 2 to 3 times more efficient (in terms of energy density) than fully reinforced designs, and are characterized by complex branching structures that route around and across the body to exploit pose-induced body deformations. Physical validation tests in 2D show our simulation model is in agreement with experimental results. We fabricated physical prototypes for the arm and the knee using a simple but easy to deploy heat-transfer vinyl (HTV) process. We found that users consistently rated our optimal designs as more resistive when compared to baselines. Additionally, we explored the performance of our method in more complex use cases, such as back posture support.

\begin{figure*}[!ht]
 \center
  \includegraphics[width=2.0\columnwidth]{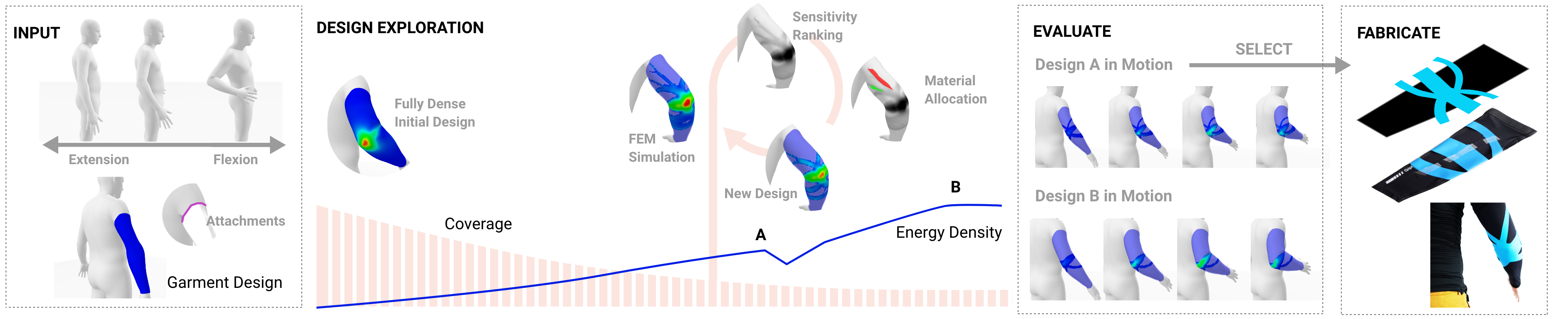}
  \caption{System overview description using the example of the arm flexion as an input motion, and the arm sleeve region as the garment to be reinforced. From left to right. \emph{Input:} A designer imports a sequence of body poses and labels the rest pose (zero energy) and final pose (optimization objective). The garments are defined on the surface of the body. \emph{Design Exploration:} The on-body topology optimization produces a sequence of optimal designs as a function of the reinforced area (budget). The background bars represent the area coverage of reinforcements. \emph{Evaluate:} Designer picks between two candidates and evaluates them in simulation over the full motion sequence. \emph{Fabricate:} the 3D on-body surface design of the reinforce regions is then flatted, cut and heat-transferred into the soft sleeve to produce the kinesthetic garment.}
  \label{fig-system} \vspace{-.25cm}
\end{figure*}

\section{Related Work}

\paragraph*{Design of Garments}
Designing garments with specific properties has been extensively investigated in graphics literature. Typical approaches consist of adjusting 2D pattern designs with regards to optimization criteria \cite{moore2018precision}, \REV{or by directly controlling textile properties through  knit and stitch patterns \cite{liu2021knitting}}. Physics-based pattern generation has been employed in multiple works, including for friction and pressure distribution \cite{montes2020computational, wang2010pattern} in skin-tight clothing, friction minimization \cite{montes2020computational}, and for automatic pattern generation from high-level user input \cite{Bartle16Physics}. \REV{Wang et al. develop a method for creating woven 2d patterns of elastic braces based on tunable springs that induce the desired normal pressure in the garment's worn state \cite{wang2007woven}}. Aesthetic criteria are considered by Kwok et al. who use an evolutionary strategy to generate a diverse set of garment designs \cite{kwok2015styling}. An emergent topic is designing garments that consider motion \cite{liu2021knitting}. These so called 4D-garments can be designed to minimize pressure from hard materials and sliding from soft-material by creating a multi-material integrated knitting map. In our work, we extend the notion of considering motion in computational garment design by providing kinesthetic feedback in response to particular motions. Similar to Montes et al. \cite{montes2020computational}, our work employs a physically based model, however, allowing for additional degrees of freedom in terms of cloth lift-off and leveraging a structural optimization approach instead of 2d pattern optimization.  






\paragraph*{Structural Optimization}
The question of how to best distribute a given amount of material such as to obtain optimal performance is a central problem in engineering \cite{bendsoe2013topology}. Topology Optimization (TO) has been employed in graphics to explore the intersection of structural objectives with aesthetic guidance by the user \cite{martinez2015structure, schumacher2016stenciling}. Recent work from Liu et al. \cite{liu2018narrow} pushed the limits in terms of grid resolution, demonstrating that, when given sufficient resolution, TO is able to recover structures similar in complexity to those found in Nature. In an elastic material setting, Skouras et al. leverage a material interpolation and penalization approach (SIMP) to optimize the material distribution of actuated characters in order to achieve a target deformation behaviour \cite{Skouras13Computational}. Bruns et al. apply the SIMP method for elastic structures undergoing large displacements \cite{bruns2001topology}, while Huang et al. demonstrated that gradient-free, evolutionary strategies such as BESO can also handle such cases \cite{huang2008topology}. Closer to our setting, structural optimization is a common approach used in cast design. Zhang and Kwok \cite{zhang2019customization} adapt SIMP into a two-manifold surface and enable efficient personalized cast designs. Personalized casts have also been optimized for thermal comfort \cite{zhang2017thermal} using an FEM in the loop strategy to selectively thicken material to increase structural stability. \REV{Topology optimization on manifolds can also be achieved by establishing a conformal map from 2D to 3D surfaces and evolving the boundary using the level-set method \cite{ye2019topology}.} While previous work has focused on topology optimization in 2D or 3D settings, we explore an on-body TO approach for automated synthesis of reinforcement structures embedded in complex-shaped 3D surfaces.


\paragraph*{Augmenting Textiles}
Kinesthetic garments belong to a wider class of textile-based compliant interfaces that are augmented with sensing and actuating systems \cite{sanchez2021textile}. Passive and quasi-passive systems have been developed to support locomotion \cite{di2020pneumatic} and provide lifting support \cite{park2019suit} using shape memory alloys. Ortiz et al. \cite{ortiz2017energy} take an optimization approach to optimally combine and place passive elastic cords, clutches, and dampers to reduce the
force and power required by a person to generate lower body motion.
Fully active systems embedded onto soft garments have been designed for  flexion/extension of the elbow joint \cite{masia2018soft}, grasping \cite{masia2018soft, hinchet2018dextres}, and for locomotion assistance \cite{kim2019reducing, lee2018autonomous}. In these works, load paths and anchoring body sites need to be carefully considered for effective transmission of forces. Sensor placement on the lower body was investigated by Gholami et al. \cite{gholami2019lower}, who use an evolutionary method to find optimal placement of fabric sensors to recover lower body kinematics. Whole-body sensing and vibrotactile actuation was used in providing feedback on back posture information \cite{barone2019sensory}. While kinesthetic garments are purely passive, they still allow for an information feedback loop that is specified at design time. They also similarly depend on the same efficient routing and anchoring to effectively transmit their forces.


\section{Overview}
\label{sec:system-overview}

Our computational pipeline supports designers throughout the complex task of designing kinesthetic garments. See Fig. \ref{fig-system} for a visual summary.

\paragraph*{Input and Parametrization} As high level input to our pipeline, designers specify the motion they wish to provide feedback to as a single rest and deformed state of the body. 

In our tool, motions are specified based on the STAR/SMPL parametric human body model \cite{osman2020star, loper2015smpl}. We use STAR for its large space of body shapes and its ability to represent the effects of pose-induced deformations such as muscle bulging. Realistic motions can be sampled from the AMASS dataset \cite{mahmood2019amass} or, alternatively, using existing methods that can recover user motion and personalized shape from RGBD sensors \cite{yu2018doublefusion}. 

Next, designers select the dimensions and placement of the soft garment onto which reinforcements will be placed (black sleeve in Fig. \ref{fig-system}). We parametrize the rest configuration of garments as 2D surfaces embedded on the body mesh, while during simulation, the garment is free to lift-off from the surface. Reinforcements are simulated on the garment mesh as a per-element material property: each element of garment material is assigned one of two materials, i.e., \textit{Cloth} or \textit{Reinforced Cloth}.

\paragraph*{Reinforcement Optimization}
We first consider the performance criteria a designer might employ, and therefore, the choice of objective function of our method. Since the overall goal is to resist a specific pose or motion, the work that the user puts in to reach that state should be maximized. In other words, the more energy stored in the reinforcements, the more energy the user has to spend getting into the specified pose. Thus, the overall energy of the kinesthetic garment in the deformed pose is a useful metric for determining how well it would resist the given motion. Comfort and weight is another factor in the design. In general, more coverage by reinforcement material can make the garment stiffer, possibly heavier, and more difficult to put on. Thus, reducing the effective coverage of reinforcements is also a useful objective for a designer. 
Our automatic design method combines these two objectives, by maximizing the energy density of the garment reinforcements, while progressively reducing the amount of garment covered by reinforcements. 

\paragraph*{Design Selection and Evaluation}
The output of our optimization stage is a continuous sequence of garment reinforcement designs, from none to full coverage, that designers can explore and analyze. Each design is the result of an FEM simulation that is based on a tailor built garment-on-body model, constrained to a particular material budget. In the evaluation stage, designs can be compared by simulating the mounted garment over the full motion sequence (that includes in-between frames). This allows designers to understand how the garment deforms, how the energy storage changes as a function of time, and to make an informed selection. 



\paragraph*{Export and Fabrication}
In the last step of the process, one final design is exported and fabricated using conventional multi-layer fabric process. In this paper, we demonstrate it via the a standard heat-transfer vinyl (HTV) application process, adding as a reinforcement material a stiffer and stretchable thin film onto the full stretchable garments (see blue lines on top of the black sleeve in output garment of Fig \ref{fig-system}).  While developing the meshes (i.e. unrolling onto a flat surface) is currently a manual task, it could be automated in a straightforward manner. 

\section{Garment-on-Body Model}
\label{sec:method-model}
In order to evaluate the performance of a given design for a specific motion, we must compute the deformations induced in the reinforced garment by given body poses. To this end, we build a computational model for kinesthetic garments that supports on-body stretching and sliding, softens under compression, and allows for lift-off separation. 
This model allows us to evaluate the performance of a given design in terms of energy density, and this metric can then be exploited by our automatic design method. In the following section, we describe each part of the model.


\subsection{Body}




As a key requirement, our computational model must ensure that the garments do not penetrate into the body and that they are able to smoothly slide over its surface.
While the STAR/SMPL parametric model is able to produce realistic body shapes, the corresponding piece-wise linear surface meshes induce gradient discontinuities for sliding motion that are problematic for continuous optimization methods.  
For this reason, we convert the discrete surface meshes from STAR/SMPL into a smooth representation based on  implicit moving least squares (IMLS)~\cite{oztireli2009feature}. The resulting smooth surfaces are implicitly defined as the zero level set of a signed distance field
\begin{equation} 
\label{eq:IMLSField}
\Phi(x) = \frac{\sum_k{\bn_k\cdot(\bx-\bv_k)\phi_k(x))}}{\ \sum_k{\phi_k(x)}} \ ,
\end{equation} 
where $\bv_k$ and $\bn_k$ are vertex positions and normals of the body, respectively, $\bx$ is a garment vertex, and $\phi$ is a locally supported kernel function,
\begin{equation}
\phi_k(x)=\left(1-\frac{\norm{\bx-\bv_k}^2}{h^2}\right)^4 \ ,
\end{equation}
that vanishes beyond the cut-off distance $h$. In practice, $h$ is pre-computed as a function of the local neighbourhood vertex distance defined by each $\bv_k$---we use twice the average edge length of its one-ring. Based on this distance field, we formulate a penalty energy that attracts garment vertices inside the body to its surface, where $\bx_i$ are the garment vertices:

\begin{equation}
\label{eq:BodyEnergy}
E_\mathrm{body}(\Phi)=\begin{cases} 
      \sum_i{\Phi(x_i)^2} & \Phi(x_i) \leq $0$ \\
      $0$ & \text{otherwise}\ .
   \end{cases}
\end{equation}

\begin{figure}[h]
 \center
  \includegraphics[width=1.0\columnwidth]{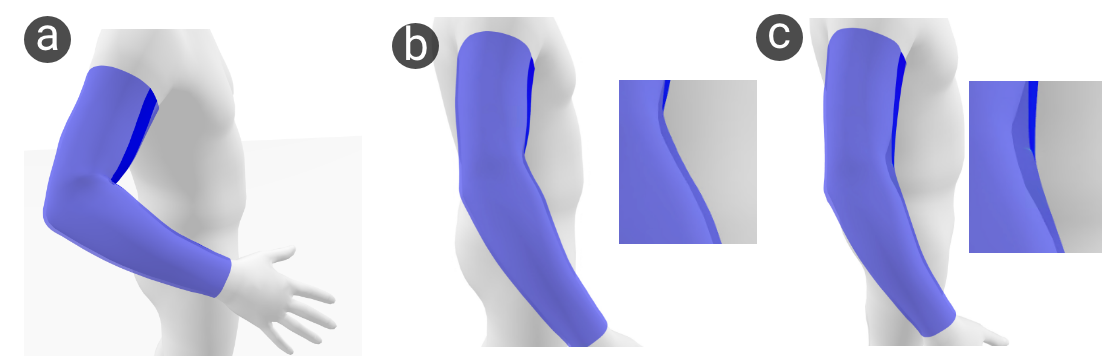}
  \caption{Starting from a neutral position (a) and moving to an extended position, we compare the behaviour between bi-lateral (b) and our uni-lateral penalty contact model (c) that prevents intersections between cloth and body while allowing lift-off. The total energy of reinforced material in (b) is 30\% higher than in (a).}
  \label{fig:lift-off} \vspace{-.25cm}
\end{figure}

Note that the above expression is a uni-lateral penalty function, i.e., it prevents the garment from penetrating into the body but allows for lift-off separation as required. 
An example where this uni-lateral formulation is essential can be in Fig \ref{fig:lift-off}, where taut material over the inner elbow lifts off under tension during arm extension. 

\subsection{Garment and Reinforcements}
We model reinforced garments as bi-material distributions where each triangle element $e$ is assigned a specific material property. \REV{The Young's modulus and Poisson ratio of the garment material are set to either \textit{cloth} or \textit{reinforced cloth} through the design variable $\materialdensity^e \in [{0, 1}]$, resulting in a per element strain energy density.}

\REV{The garments are discretized as a triangle mesh with sets of nodal positions $\mathbf{x}=(\mathbf{x}_1,\ldots,\mathbf{x}_m)\in\mathbb{R}^{3n}$ and $\bar{\mathbf{x}}=(\bar{\mathbf{x}}_1,\ldots,\bar{\mathbf{x}}_m)\in\mathbb{R}^{3n}$} for deformed and undeformed poses, respectively. 

To quantify the deformation for a given triangle, we compute the deformation gradient $\bF_C$ that maps edge vectors $\bar{\be}_{i,j}$ from the undeformed triangle to their deformed counterparts $\be_{i,j}$, i.e.,
\begin{equation}
    \be_{i,j}=\bF_C\bar{\be}_{i,j} \ .
\end{equation}
We note that computing $\bF_C$ in this way is equivalent to a standard finite element discretization with linear triangle elements. 

Based on the per-triangle deformation gradient $\bF_C$, we compute the corresponding right Cauchy-Green tensor $\bC_C$ ,
\begin{equation}
\label{eq:FandC_Cloth}
\bF_C=\frac{\partial \bx}{\partial \bar{\bx}} \quad \text{and} \quad \bC_C = \bF_C^T\bF_C\ .
\end{equation}

Since $x\in\mathbb{R}^3$ and $\bar{x}\in\mathbb{R}^2$, the deformation gradient $\bF_C$ is a $3\times2$-matrix and $\bC_C$ is a $2\times2$-matrix describing the deformation of the  element with respect to rest state coordinates. 

Based on $\bC$, we define an elastic energy density based on the classical compressible neo-Hookean model \cite{bonet_wood_2008}:

\begin{equation}
\label{eq:energyDensityGarment}
    \energydensity_\mathrm{garment}^e(\bC, \materialdensity^e) = \frac{\mu}{2}(I_C-2) - \mu\log{J}+\frac{\lambda}{2}(\log{J})^2 \ ,
\end{equation}

where $I_C=\text{tr}(\mathbf{C})$ is the first invariant of the right Cauchy-Green tensor that captures all deformations, and $J=\text{det}(\bF_C)$ is the relative area change. The material parameters \REV{$\lambda$ and $\mu$ correspond to Lame's first parameter and shear modulus, which can be converted to Young's modulus and Poisson's ratio.}

\begin{figure}[h]
 \center
  \includegraphics[width=1.0\columnwidth]{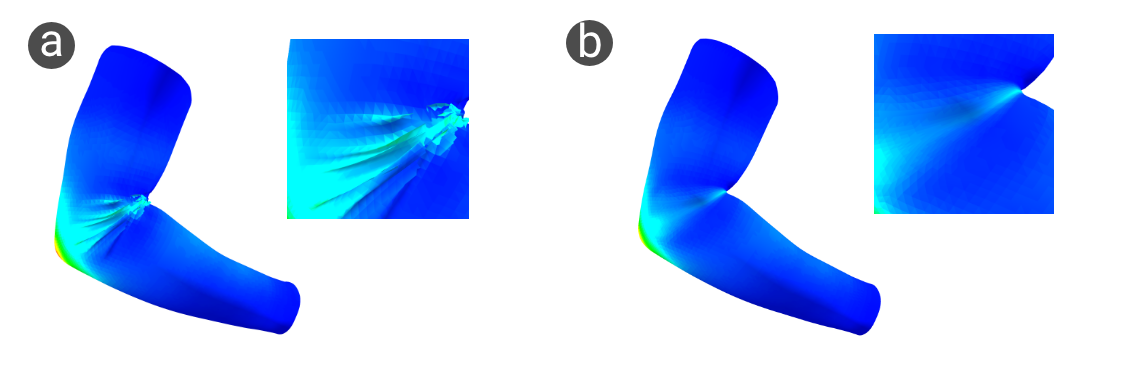}
  \caption{Wrinkles that emerge in response to compression can generate spurious deformation patterns and geometric artefacts (a). With a relaxed energy formulation that does not penalize compression (b), deformations are smooth and free from compression artefacts.}
  \label{fig:wrinkling} \vspace{-.25cm}
\end{figure}

In a setting of fully constraining the CST elements to a manifold (i.e. mesh surface), this energy would result in garments that resist compression, a property that cloth garments do not exhibit. Prior work \cite{montes2020computational, skouras2014designing} deals with this problem by defining a relaxed strain energy \cite{pipkin1986relaxed} for compressive modes of the respective materials. Our formulation allows garments to lift off the surface of the body (see \ref{eq:IMLSField}) and thus produce wrinkles in areas that experience compressive forces. However, without sufficiently high mesh resolution and proper handling of self contacts, these wrinkles tend to create spurious deformation patterns (see \ref{fig:wrinkling}) that can mislead the optimization method.  We therefore adopt a mixed approach that does not penalize compression but allows cloth to lift off the body surface, e.g., in response to stretching over concave regions (see Fig. \ref{fig-arm-flex}). 

Since the energy density is constant over each triangle element as per its definition, we can integrate this quantity into a final per element energy and sum over these to arrive at the final garment energy:

\begin{equation}
\label{eq:energyGarment}
E_{\mathrm{garment}}(C, \materialdensity) = \sum_e t^e  A^e\energydensity_\mathrm{garment}^e(C, \materialdensity^e)\ ,
\end{equation}
where $A^e$ is the per-triangle area in the undeformed configuration and $t^e$ is the per element thickness of the fabric.

\subsection{Attachments \REV{and Contact}}
Kinesthetic garments typically have straps or elastics at their boundaries that prevent excessive sliding. \REV{Therefore, we specify Dirichlet conditions at the attachment points that we enforce through soft constraints.} We allow the designer to specify areas of the garment that would be attached to the body, for example, via belts, velcro loops, or using sticky materials such as silicone. Typically, these might be placed at the boundaries of the garment. 

Similar to $\materialdensity^e$, for each triangle we specify whether it is attached or not. To avoid creating overly stiff boundary elements, we use the triangle center as the origin rather than its vertices. We formulate this coupling potential as a zero-length spring:

\begin{equation}
\label{eq:energyAttachments}
E_\mathrm{attachments} = \frac{1}{2}k(x^c-x^c_0)^T(x^c-x^c_0) \ ,
\end{equation}
where $k$ is the spring stiffness, and $x^c$ and $x^c_0$ denote the deformed and undeformed positions corresponding to the centers of attached elements. \REV{We choose a relatively low value of $k=0.002$ (compared to the base fabric) to discourage designs that do rely on `pulling on the boundary`.}

\REV{Attachments ensure contact between the garment and the body. Contact between body and garment may create an additional stimulus, but it is not required for leveraging strain energy for feedback. While our formulation allows lift-off, some part of the garment is always in contact with the body, either through a loop structure or an attachment point.}

\subsection{Simulation}
\label{sec:method-simulation}
With the model and energies defined above, we perform a quasi-static simulation by solving the following unconstrained optimization problem:

\begin{equation}
\label{eq:CoupledSystemEnergy}
\bx^* = \arg\min_{\bx}  \quad E_\mathrm{body}(v, \bx) + E_\mathrm{garment}(\bx, \materialdensity) + E_\mathrm{attachments}(\bx)  \ .
\end{equation}

To ensure that $\bx^*$ is in a state of equilibrium, the minimization must ensure that derivative of the above system vanishes with respect it's degrees of freedom $\bx$ in the deformed configuration:

\REV{
\begin{align}
	\label{eq:forceBalance}
	\bff(x^*) & = \frac{\partial E_\mathrm{body}}{\partial x} + \frac{\partial E_\mathrm{garment}}{\partial x} + \frac{\partial E_\mathrm{attachments}}{\partial x}= \mathbf{0}\ .
\end{align}
}

To solve this minimization problem, we use the L-BFGS \cite{liu1989limited} optimizer provided by PyTorch \cite{pytorch} which affords us with super-linear convergence, while also allowing us to simulate fairly dense topologies. We consider simulations converged once their gradient norm reaches 1e-7. The simulation is performed on a PC running a quad core 3.6Ghz I9 CPU and an NVidia 2080TI GPU. All calculations are done on the GPU. 

\subsubsection{Initialization}
\label{sec:model-initialization}

Since the garment is parametrized as a subdivided subset of the body mesh, we can initialize the garment positions in rest and deformed states in a straightforward manner by simply copying the position of the body vertices to the corresponding garment vertices. This initialization is important for our local IMLS field to produce meaningful gradients, as any garment vertices that are initialized too far inside the body would have no energy to push them to the surface.

\section{On-Body Topology Optimization}
\label{sec:design-method-selection}
In order to automatically generate energy-dense designs for a given motion and garment, we look towards topology optimization (TPO). The aim of such methods is to find an optimal \textit{material distribution} within a prescribed domain under load, minimizing for a particular cost function (i.e. compliance) and staying within a material budget \cite{bendsoe2013topology}. We surveyed two widely used TPO continuous and discrete approaches. SIMP is a continuous method that penalizes intermediate material densities with the power-law to encourage near-binary material distributions \cite{sigmund200199}. BESO \cite{huang2007convergent} is a bi-directional evolutionary method that iteratively refines a binary material density $\{0, 1\}$  distribution towards a given material budget. In penalization based approaches like SIMP, intermediate results contain invalid non-binary material parameters and are thus not usable. Intermediate results in BESO on the other hand are valid as they are always in the $\{0, 1\}$ set. Comparisons between SIMP and BESO have been undertaken and show that they can produce similar results \cite{huang2010further}, however, only BESO can create a continuous space of optimal designs in one shot. The BESO algorithm is a better fit for our design scenario, allowing a continuous and coherent \textit{exploration} of the design space of possible reinforcement layouts.

\subsection{Design Objective}
\label{sec:design-objectives}
As a basis of our algorithm, we use the multi-material version of BESO proposed by \cite{huang2009bi}. In the standard setting, elements are entirely added or removed from the domain, while in the multi-material setting, removed elements are set to a weaker material. These two approaches are known as hard-kill and soft-kill BESO respectively \cite{ghabraie2015improved}. 

As noted in Section \ref{sec:system-overview}, our goal is to increase the energy density of the garment while gradually decreasing the surface area instrumented with reinforced material. We define the following optimization problem:

\begin{equation}
\label{eq:BESOObjective}
\begin{split}
\bd^* = \arg\max_{\bd}  \quad E_\mathrm{garment}(x^*, \bd) \\ 
\textrm{s.t.} \quad \sum_e{A^e} d^e=A^*\ ,\quad \REV{\bff(x^*)= \mathbf{0}}
\end{split}
\end{equation}

where $A^*$ is the prescribed area of the domain, and $\bd$ is the design vector determining the material assignment which, for each element, can take on the values $d^e \in {0, 1}$. Note that when $\materialdensity^e$ is $0$, its area does not count toward the target material area. In effect, for a given target area $A^*$, we are looking for the optimal material assignment $\bd^*$ that maximizes the energy of the garment in its equilibrium state $\bx^*$. 

\subsection{Sensitivity to Objective}
In the BESO method, the optimal topology $\materialdensity^*$ is determined according to a relative ranking of sensitivity numbers, where the sensitivity of each element in the domain is evaluated w.r.t. to the objective function. The sensitivity number of linear materials has been derived simply as the per-element strain energy $E^e$ \cite{huang2007convergent}, that is, the \textit{variation} in the total strain energy is the same as the strain energy element if it were to be added or deleted. To take into account our non-regular mesh topology, we  modify this by dividing by the element area. Thus, in our case, the per-element sensitivity $\sensitivity^e$ is equal to the elemental strain energy density $\energydensity_\mathrm{garment}^e$.

In the multi-material setting, sensitivity numbers need to be modified to take into account the elastic moduli of each material. We follow a similar approach as Huang et al. \cite{huang2009bi}, integrating the concept of penalization and material interpolation from SIMP, resulting in the following formulation for two materials:

\begin{equation}
\label{eq:sensitivty}
\sensitivity^e=\begin{cases} 
      \frac{1}{2} \left[ 1-\frac{E_2}{E_1} \right] \energydensity_\mathrm{garment}^e & \text{material 1 (Reinforced Cloth)} \\
      \frac{1}{2} \frac{d_\mathrm{min}^{p-1} \left( E_1 - E_2\right)}{d_\mathrm{min}^{p} E_1 + \left(1- d_\mathrm{min}^{p} \right) E_2}  \energydensity_\mathrm{garment}^e & \text{material 2 (Cloth)} \ .
   \end{cases}
\end{equation}

where $E_1$ and $E_2$ correspond to the Young's modulus of reinforced cloth and cloth respectively, and $d_\mathrm{min}$ is the minimum value for the design variable $\materialdensity$. While in our case $\materialdensity$ is in the set ${\set[d \in {0, 1}]}$, setting $d_\mathrm{min}$ to $0$ would simply switch the method to the hard-kill version as the secondary material would have $0$ sensitivity. The power exponent $p$ interpolates between the influence of the first and second material. For all experiments, we set $d_\mathrm{min}=0.001$ and $p=1.6$.


\subsection{Smoothing and Filtering}
Performing topology optimization with constant strain elements can result in the well known problem of \textit{checkerboarding}, where the sensitivity of elements can become discontinuous across boundaries.
To overcome this problem, a filter is applied to both smooth the sensitivity across elements, and to interpolate sensitivity numbers between the boundaries of multiple materials. The filter works by transferring elemental sensitivities into their connected nodes:

\begin{equation}
\label{eq:BESOfiltering}
\tilde{\sensitivity_i} =  \frac{\sum_{e \in \epsilon} A^e \sensitivity^e }{\sum_{e \in \epsilon} A^e} \ .
\end{equation}

where $\epsilon$ is the set of elements connected to node $i$, $A^e$ and $\sensitivity^e$ are the element area and sensitivity respectively. It is important to note that nodal sensitivities carry no physical meaning, but can be interpreted as nodal sensitivity \textit{density}. We use simple averaging $\tilde{\sensitivity^e} = \sum_{i \in e}^N \tilde{\sensitivity_i} / N $ to redistribute sensitivity values back onto elements. The sensitivity numbers are further temporally smoothed with their historical information:

\begin{equation}
\label{eq:BESOfiltering}
\tilde{\sensitivity_i} = \frac{1}{2} \left(  \tilde{\sensitivity_{i,k}} + \tilde{\sensitivity_{i,k-1}} \right) \ .
\end{equation}

where k corresponds to the current iteration of the algorithm.

\subsection{BESO Procedure}
\label{sec:BESOprocedure}


With the elemental sensitivities $\tilde{\sensitivity^e}$ in hand, we can apply the BESO structural refinement procedure. First, The design vector $\materialdensity$ is initialized with a fully dense structure resulting in a starting area $A_0$. The following steps are repeated until convergence: 

\begin{enumerate}
  \item Finite element analysis is performed bringing the garment mesh into an equilibrium state $x^*$, resulting in sensitivity numbers $\sensitivity^e$. 
  \item Apply spatial and temporal filtering to obtain $\tilde{\sensitivity^e}$
  \item Rank sensitivity numbers $\tilde{\sensitivity^e}$ of all elements in the domain, and find the threshold $\sensitivity_{th}$ such that the area of all elements with higher sensitivities satisfies the current iteration's target area $A_i$.
  \item Update the design variable $\materialdensity^e$ for all elements with sensitivities below  $\sensitivity_{th}$ to $0$, and set all those above to $1$
  \item To stabilize the procedure , a ratio $AR_\mathrm{max}$ controls the upper limit on elements that can be switched from $0$ to $1$. 
  \item Update the target area for the next iteration $A_{i+1} = A_i(1 \pm ER)$, where $ER$ is the evolutionary ratio. If $A_i \equiv A^*$, then set $ER$ to $0$.
\end{enumerate}

We consider the optimization converged when the target $A^*$ is reached and the change in energy in Eq.\ref{eq:BESOObjective} is less than a predefined threshold $\tau$ over $N$ previous iterations. \REV{A detailed procedure of two material BESO can be found in \cite{huang2009bi}.}



\section{Results}\label{sec:results}
We conduct a three-part evaluation of the proposed computational pipeline. We start by validating our simulation model against a set of physical cloth samples in 2D. Next, we explore the capabilities of our method in a \REV{simulated} on-body setting, generating designs for a diverse set of motion and garment pairs. We then fabricate and physically test 2 of these designs in a blind comparative user study.

\subsection{Material Parameters and 2D Validation}
\begin{figure}[t]
 \center
  \includegraphics[width=1.0\columnwidth]{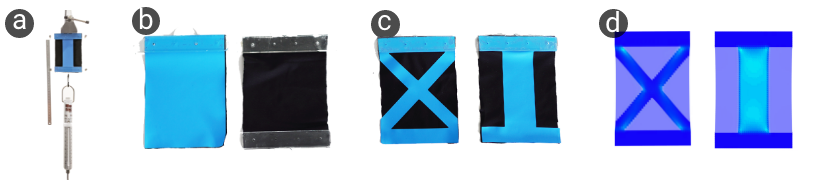}
  \caption{Material stiffens and 2D model validation. a) Built pull-tester. b) Samples of \textit{cloth} and \textit{reinforced cloth} used for Young's modulus to recover ($0.5$ MPa and $5.7$ MPa, respectively). c) A separate set of samples was fabricated and tested against d) simulated designs within an agreement of 10\%.}
  \label{fig-2d-experiment} \vspace{-.25cm}
\end{figure}

We first conducted an experimental characterization of the material parameters of \textit{cloth} and \textit{reinforced cloth}. For this purpose, we custom built a pull-testing apparatus (See fig. \ref{fig-2d-experiment}) and measured the uni-axial forces resulting from a controlled in-plane displacement. For each material, we fabricated a $10\times14$ cm of fabric and clamp it at its two ends. This gives a $10\times10$ cm patch of active deformable area. Tensile forces were measured on a Pesola spring-scale. Fabric samples were taken directly from the arm sleeves used in the fabrication section. 
Based on force-displacement measurements, we determined a Young's modulus of $0.5$ MPa for \textit{cloth} $5.7$ MPa for \textit{reinforced cloth}. The thicknesses of the base cloth was $0.27$ mm and $0.35$ with reinforcements added. A Poisson's ratio of $0.33$ was used based on prior work on elastic knitted fabrics \cite{jinyun2010poisson}.

Using the values derived above, we constructed an identical virtual testing apparatus and perform simulation studies in 2D. We compare the force output of two virtual samples against their physical counterparts, 
each with 40\% area covered by reinforcements (See Fig \ref{fig-2d-experiment}). 
Our results show that the simulated and physical behaviour are in agreement for this validation setup. For example, at 10\% induced deformation in both samples, for a) the \textit{Line}: we measured $5.5$ N with the experiment and obtain $5.4$ in simulation (2\% difference); while for b) the \textit{X} design we measured $10.8$ N against the $9.9$ N of simulations (8\% difference).


\subsection{\REV{Simulated} On-body Designs}
\label{sec:eval-on-body-designs}
\begin{figure}[t]
 \center
  \includegraphics[width=1.0\columnwidth]{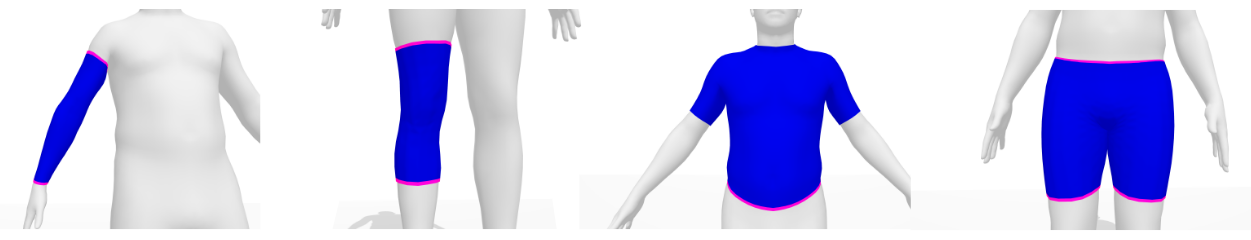}
  \caption{A range of garments designed with our tool. Pink areas represent attached areas. All garments start in a fully reinforced setting, which is represented with opaque blue color.}
  \label{fig-garments} \vspace{-.25cm}
\end{figure}

The goal of this study is to demonstrate both the diversity of designs as well as their performance in terms of energy per unit area. We begin by sampling the required poses and motions from the AMASS dataset \cite{mahmood2019amass}. We use only the pose parameters and zero out the $\beta$ (shape) parameters. Four garments are designed using our tool to cover a variety of body sites (See Fig. \ref{fig-garments}). We set the following BESO parameters for all experiments: ER = 1.5\%, $AR_{max}$   = 1.5\%, $A^*$ = 15\%, except in the case of knee-flexion and slouching, where $A^*$ = 10\%. As a primary measure, we use the objective function defined in Eq. \ref{eq:BESOObjective}, that is, the energy density of just the reinforced part of the garment. We normalize this value to the fully dense design (equal to the first iteration of the algorithm) to allow comparison of designs in terms of their relative energy density. All designs are evaluated in simulation, and a subset of these designs are fabricated and evaluated by users. \REV{Parameters and performance for all experiments are listed in Table~\ref{tbl:performance}.}

\begin{table}[h]
\begin{tabular}{ p{2cm} p{0.75cm} p{0.75cm} p{1.0cm} p{2.0cm}  }
\hline
model   & vertices & BESO steps & time / iter.  & energy density increase   \\ \hline
\rule{0pt}{3ex}Arm Flexion* & 7.1k & 150 & 10.65s  & 1.97x \\
Arm Extension & 7.1k & 180 & 10.88s  & 2.26x \\
Knee Flexion* & 3.2k & 180 & 10.47s  & 2.16x \\
Knee Extension & 3.2k & 150 & 10.24s  & 2.91x \\
Slouch* & 28.3k & 200 & 15.80s & 2.15x \\
Crouch & 9.5k & 150 & 11.70s & 2.32x \\
Lunge & 9.5k & 180 & 11.55s & 3.02x \\
\hline
\end{tabular}
\caption{Summary of parameters and performance for simulated results. * Indicates designs that were fabricated.}
\label{tbl:performance}
\end{table}

\begin{figure}[h]
 \center
  \includegraphics[width=1.0\columnwidth]{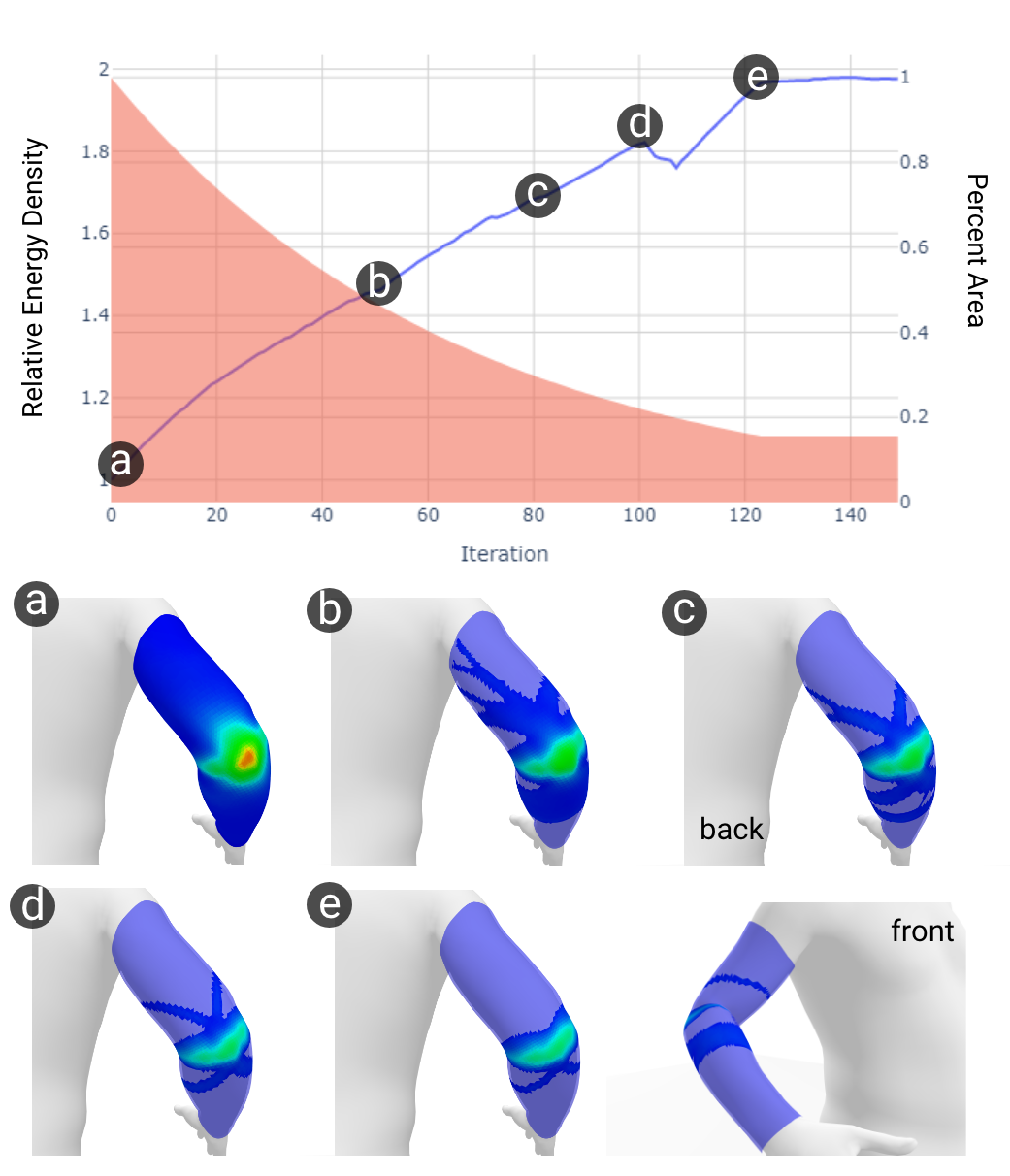}
  \caption{Optimization of Arm Sleeve targeting Arm Flexion. Starting with a fully dense design (a), the target area is reduced and reinforcements are placed in such a way as to smoothly increase the energy density of the garment. Many efficient designs can be sampled from this space (b-e). }
  \label{fig-arm-flex} \vspace{-.25cm}
\end{figure}

\begin{figure}[h]
 \center
  \includegraphics[width=1.0\columnwidth]{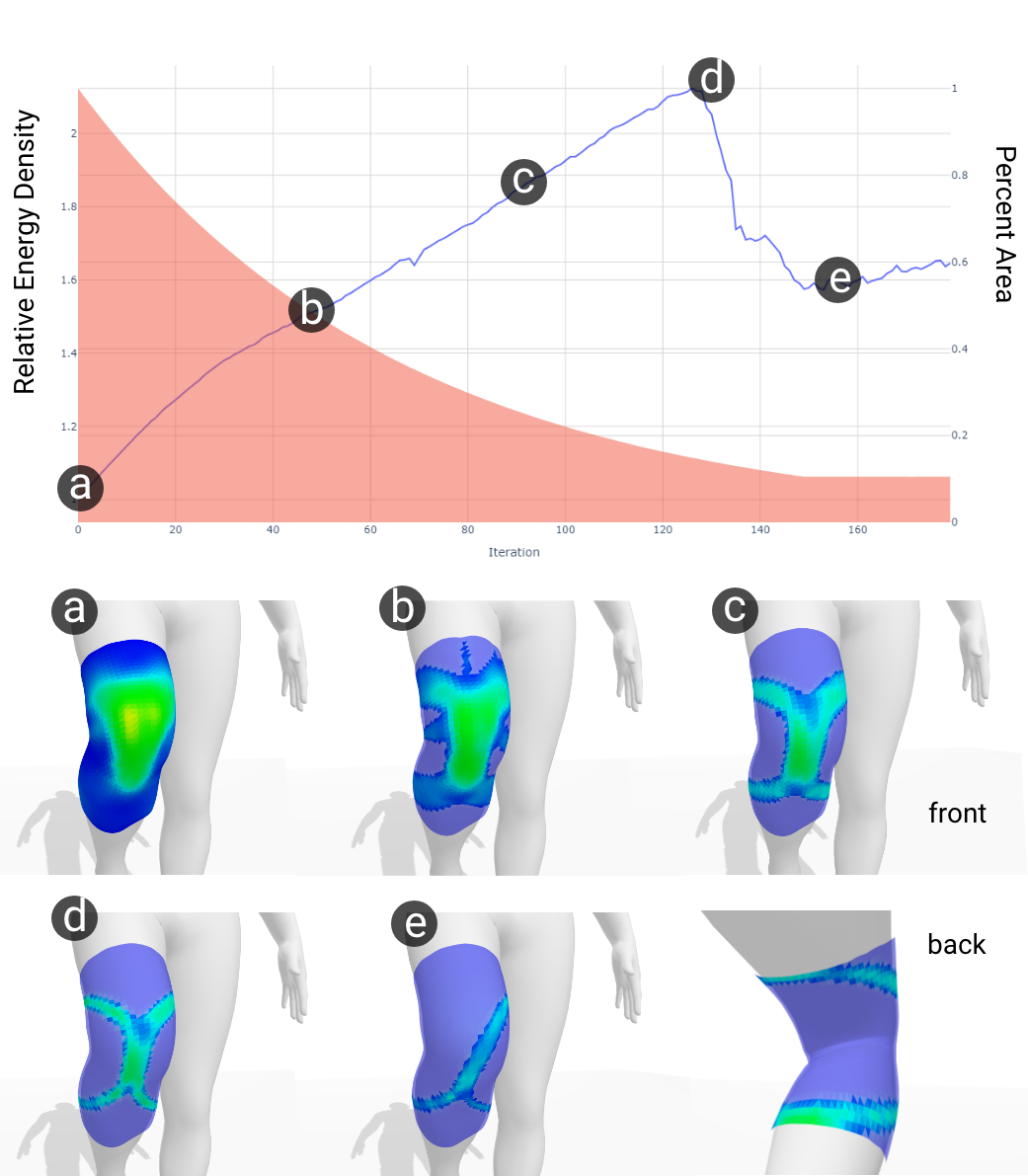}
  \caption{Optimization of Knee Sleeve targeting Knee Flexion. The energy density increases up to a point (a-d), but falls off drastically if too much material is trimmed. While the design in (e) is still more energy dense than (a), there is not enough material budget to sustain the more energy dense design in (d). }
  \label{fig-knee-flex} \vspace{-.25cm}
\end{figure}

\paragraph*{Simple motions}
We begin by studying low degree-of-freedom joints at the elbow and knee. Results for flexion of these joints is shown in Fig \ref{fig-arm-flex} and \ref{fig-knee-flex}. In the \textit{Arm Flexion} case, we see a smooth and monotonically increasing objective function, with a smoothly decreasing area profile. When target area is reached, the designs and objective function are stable. A number of designs can be sampled from this space, depending on user needs. 

In the \textit{Knee Flexion} case, we intentionally specify a smaller area budget (10\%), resulting in the sub-optimal design shown in (e). This is due to insufficient material to sustain the more energy dense X design found in (d). In this case, our tool allows designers to simply revert to the more energy dense design with slightly more area coverage.  
In both optimization routines, the energy density of designs increases by over two times. In both designs, we observe the formation of \textit{loops}---key structural features that emerge when attachments cannot be relied upon to hold the structure in place. 
One difference between the two most optimal designs in (d), is the lack of a loop directly over the joint. As the optimization is exploiting muscle bulging, there is less of this effect on the knee than in the arm.

\begin{figure}[h!]
 \center
  \includegraphics[width=1.0\columnwidth]{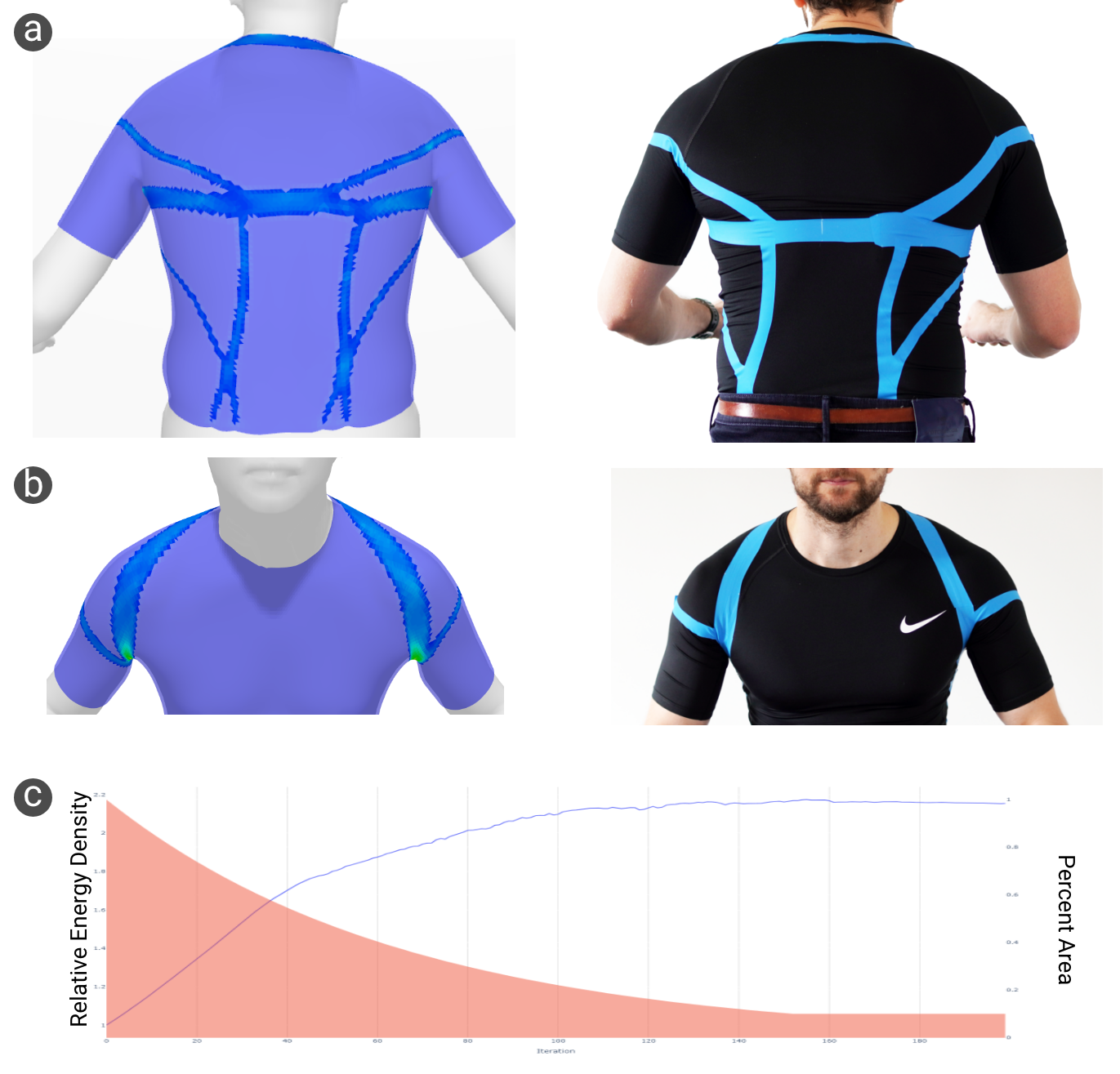}
  \caption{Optimized posture correcting shirt viewed from the back (a) and front (b). Area coverage is 10\%. Note the smooth increase in energy density (c) over the optimization process.}
  \label{fig-posture} \vspace{-.25cm}
\end{figure}

\begin{figure}[h!]
 \center
  \includegraphics[width=1.0\columnwidth]{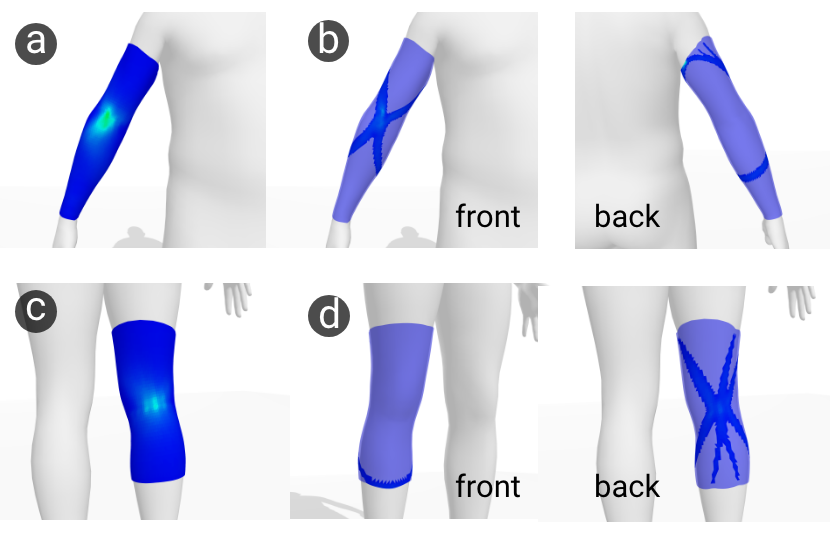}
  \caption{Knee and Arm Sleeves for Knee and Arm Extension. Fully dense designs are shown in (a) and (c), 15\% in (b) and (d). Loops are present in both designs and a spider-like web is apparent on the knee-design.}
  \label{fig-extension} \vspace{-.25cm}
\end{figure}

Next, we look at designs for the same joints and garments in the case of arm extension and knee hyper-extension. Results for these motions are shown in Fig \ref{fig-extension}. Knee-Hyper extension, where the knee joints becomes locked is not a desirable pose and can negatively affect posture. In this case, our optimization discovers two interesting features: for knee-hyper extension, a web-like structure emerges with a loop on the front of the knee. In the arm extension case, a two-sided loop wraps around to the back of the arm.

\paragraph*{Complex Motions}
We evaluated our method on motions that involve multiple joints in both the upper and lower body. For the upper body, correcting bad posture induced by slouching is one possible application of kinesthetic garments. Fig \ref{fig-posture} shows the optimization result for counteracting such a motion. The optimized design exhibits a harness-like topology with complex network of branches that connect to the base. Multiple loops are present, extending over the arms, around the front of the shoulders, and under the armpits. It is worth noting that the slight asymmetry observed in this design is due to the person-specific slouching pose.

\begin{figure}[h]
 \center
  \includegraphics[width=1.0\columnwidth]{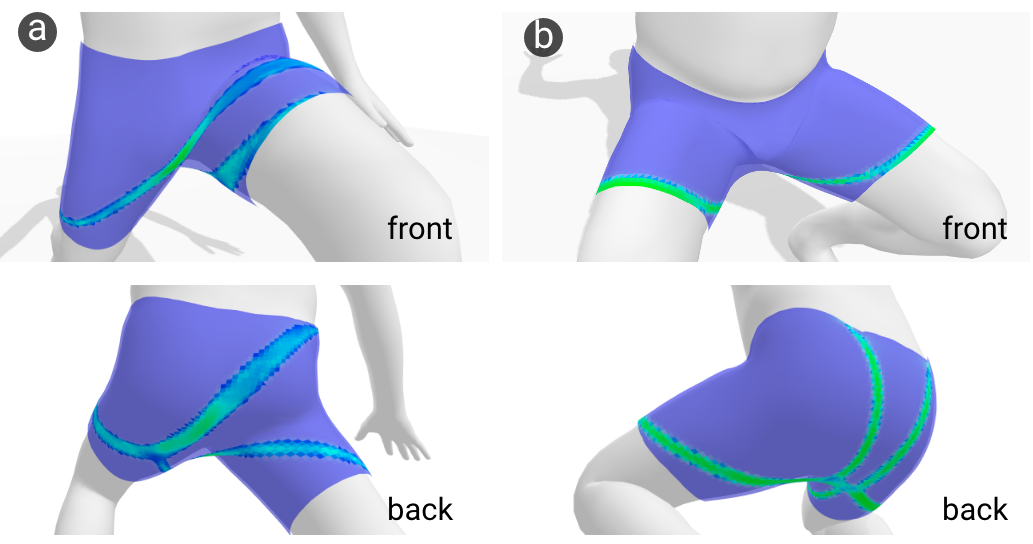}
  \caption{\REV{Optimized exercise cycling shorts for lunging (a) and crouching (b). Area coverage is 15\%.}}
  \label{fig-exercise} \vspace{-.25cm}
\end{figure}

\REV{For the lower body, we show the effect that different motions can have on a garment design. In this case, a potential application are cycling shorts designed for resisting motions during exercise. The resulting designs shown in Fig \ref{fig-exercise} are both appropriate and unique to the motion under which they were optimized. }

\begin{figure}[h]
 \center
  \includegraphics[width=1.0\columnwidth]{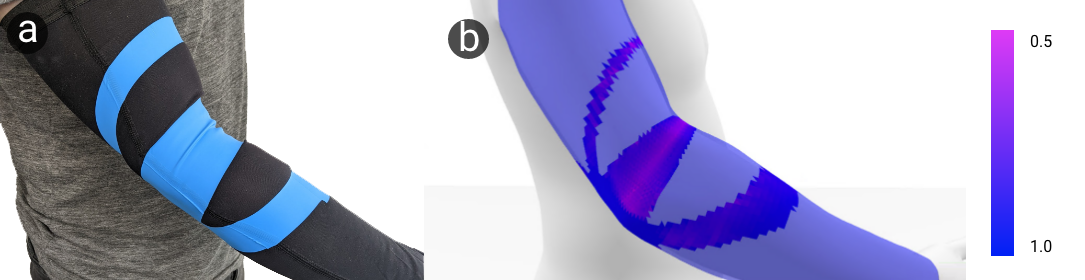}
  \caption{\REV{Wrinkling in simulation (a), compared to wrinkling in fabricated garment (b). Compression field is visualized from (blue) no compression, to (purple) compression in the transverse direction of strain.} }
  \label{fig-wrinklin-result} \vspace{-.25cm}
\end{figure}

\paragraph*{Wrinkling}
\REV{Thin sheet materials generally wrinkle at the onset of compression. Resistance to compression is therefore orders of magnitude weaker than for stretching. Our relaxed energy formulation captures this behavior at the macro level for both garment and reinforcement materials without the need for modeling fine-scale wrinkles. Importantly, it tolerates compression in a given direction while accurately capturing the more vital stretching in the transverse direction—a combination that arises frequently in our physical samples. Figure \ref{fig-wrinklin-result} demonstrates a close match between our simulated compression field and the wrinkles in the fabricated garment.}

\REV{\section{User Evaluation}}

\label{sec:eval-user}
In order to validate the effectiveness of automatically produced designs, we select a subset of motions \textit{Arm Flexion and Knee Flexion} and fabricate one optimal design for each. These designs are compared to 2 baselines: an X shape and a Line in a simple head-to-head user study. Each baseline was designed to be exactly the same area as the optimal design. \REV{We chose a line as it forms a geodesic between attachment points, while the X forms a loop on the arm.}

\begin{figure}[h]
 \center
  \includegraphics[width=1.0\columnwidth]{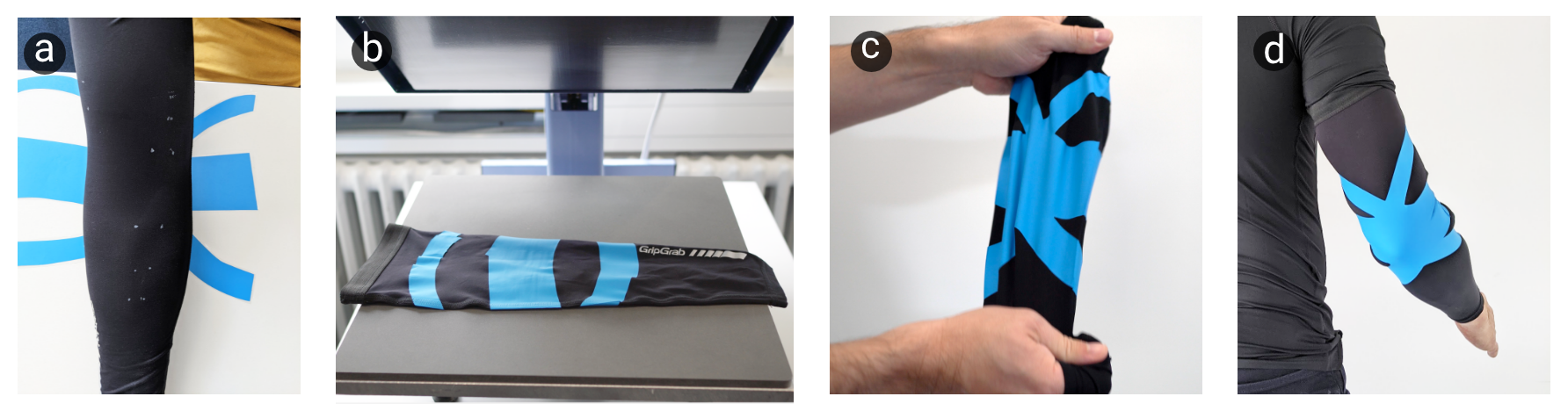}
  \caption{\REV{Garments reinforcement material (HTV) is cut (a) and transfered onto the base garment using a heat press (b). The resulting garment has high compliance (c) and can be worn as a typical sleeve (d). } }
  \label{fig-fabrication} \vspace{-.25cm}
\end{figure}

\begin{figure}[h]
 \center
  \includegraphics[width=1.0\columnwidth]{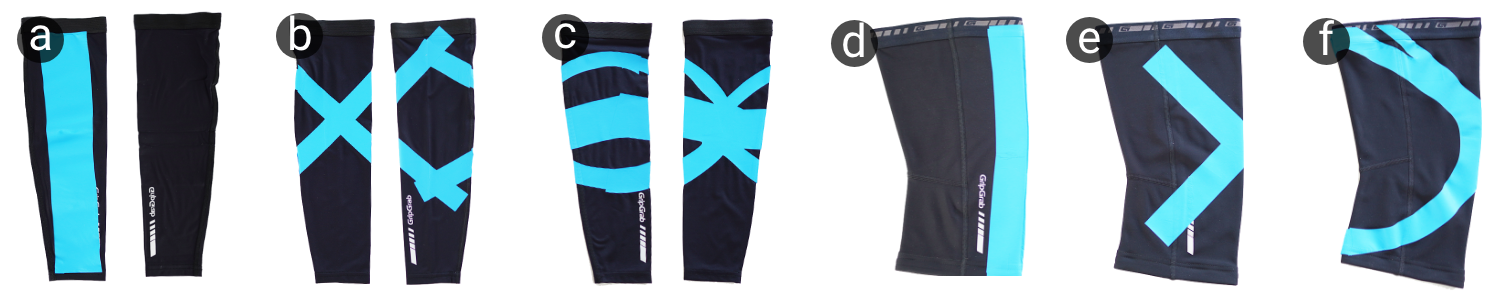}
  \caption{Fabricated Garments for user study. Line, X, and Optimal designs for elbow sleeves (a-c), and knee sleeves (d-e) respectively.}
  \label{fig-study-garments} \vspace{-.25cm}
\end{figure}

\paragraph*{Fabrication}
We fabricated 3 designs for each joint, as shown in Fig \ref{fig-study-garments}. Optimal designs are exported as meshes, developed onto a flat surface, and processed in 2D to smooth edges. We maintain the area and shape of each design throughout the process. As our material, we used the Siser EasyWeed Stretch as it is one of few vinyls that is stretchable and can recover its form after deformation. For the stretchable garments for the arm and knee we used GripGrab UV sleeves. Designs are cut with a Siluhette Cameo 4 Pro plotter and applied with a Transmatic TMH 28 heat press for 16 seconds at 160 C. Designs are weeded in reverse fashion (peeling off the material to be applied), and all material is placed on the garment simultaneously, allowing for more accurate seam placement. The fabrication process is shown in Figure \ref{fig-fabrication}.

\begin{figure}[h]
 \center
  \includegraphics[width=1.0\columnwidth]{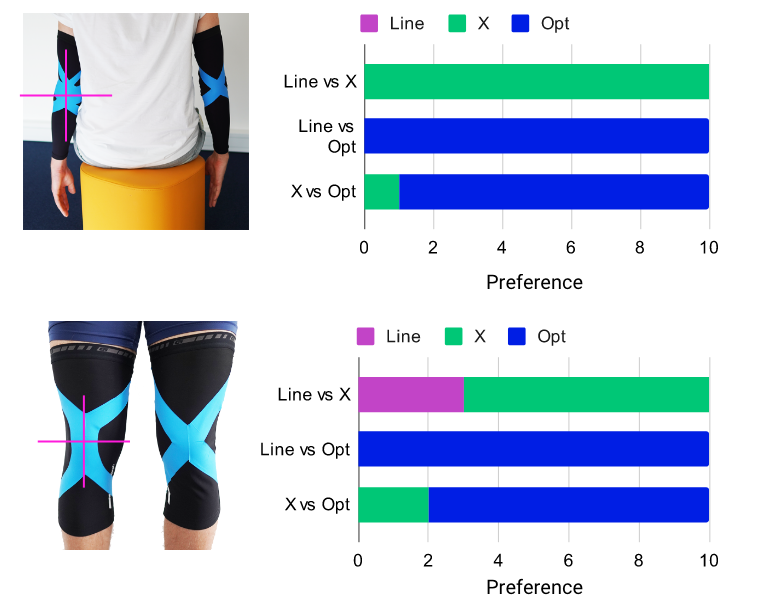}
  \caption{User study setup and preferences for Arm Flexion and Knee Flexion. The center of garments (represented with pink cross) was aligned to the mid point of the elbow/knee. Users prefer the BESO optimized design a 90\% of the times in the arm (top) and 80\% for the knee (bottom).}
  \label{fig-user-study} \vspace{-.25cm}
\end{figure}

\paragraph*{Procedure}
Five healthy adult subjects ($M$=27.6; $SD$=4.14;) were recruited. Since we only fabricated one size of our designs, participants were all male and similar in size to the template SMPL mesh. All participants wore a blindfold so they could not identify which sample they were wearing. 
For each trial, two different designs were mounted on each arm. The center of the garment designs were carefully aligned to the center of the elbow (See fig \ref{fig-user-study}.
Participants were then asked to rotate their elbow to between 45-60 degrees slowly, to get a sensation of how much the resistance each garment on their arms produces. No time limit was set, and multiple attempts were allowed.  
After this, participants gave their preference for which garment produced more resistance to their motion. Each comparison was repeated with the arms switched, so the effect of arm strength could be balanced. The order in which garments was worn was randomized. A total of 12 trials were completed for the Arm Sleeve comparisons, and the procedure was repeated in identical fashion for the Knee Sleeve.

\paragraph*{Results}
For \textit{Arm Flexion}, participants ranked the optimal design as providing more resistance in all but one case. The line design was never selected once. It's worth noting that both the X design and the optimal design contain loops in the upper and lower part of the arm, however, only the optimal design contains the loop in the center of the elbow. This could be one reason for performing better in this comparison.


For \textit {Knee Flexion}, the X design was chosen over the optimal design in two out of ten comparison, and the Line design was selected three times over the X. Participants noted several times that the Knee Flexion comparison was more difficult to make. \REV{The less consistent ranking for the knee example could be due to the comparatively strong leg muscles, which might reduce sensitivity to changes in garment stiffness.} One participant noted that he felt pressure on the front of his leg with the optimal design.

\REV{\section{Limitations}}
\REV{A limitation of our method is that it does not account for the effect on nearby motions that should not be resisted. In particular, garments that cover areas with high mobility (i.e. around the torso) could impact freedom of movement. In these cases, the objective function could be modified with additional terms penalizing energy density in such nearby motions. To generalize the method further, multiple possible motions can be taken into account where each motion can be resisted or kept free.} 

\REV{A limitation of our model is the simplified attachment and contact potential, in particular, it cannot account for stick-slip behaviour that occurs when the strain energy of the garment causes the attachments to slip from their original mounting positions. Attachments must also be specified manually by the designer which is appropriate only for pre-existing garment substrates (e.g. sleeves). Automatic placement of attachments could be integrated into the method via rope caging \cite{kwok2016rope} to enable design of custom garments.}

\REV{With regards to comfort, we assume that minimizing coverage will reduce the overall impact on the garments' original behavior. While coverage is not a direct measure of comfort, even more sophisticated functions of comfort will often involve some degree of coverage reduction \cite{zhang2017thermal}. A quantitative measure of comfort such as dense pressure can be integrated into the method to produce comfort aware designs.}

\section{Conclusions and Future Work}
\label{sec:discussion-limitations}
Designing kinesthetic garments that provide force feedback to user-defined motions is a challenging task.
In this work, we presented a topology optimization approach to this problem that automatically generates optimal designs that maximally resist specified motions for given reinforcement budgets. Through a sequence of simulation examples, we show that our method generalizes well to a large space of human motion, ranging from simple flexion and extension examples for single joints to complex motions spanning multiple joints.
The effectiveness of our designs as indicated by simulation results were substantiated by a user study with physical prototypes, showing a clear preference for our optimized design.

\REV{Although our method can generate a range of designs that users can choose from, they may want additional controls to influence the design of structures in desired directions, for example, to avoid injured areas. Such user control can be implemented via painting weight fields on the mesh, similar to \cite{schumacher2016stenciling}.}


While kinesthetic garments are passive in nature, the basic principles of maximizing energy density while considering body deformation also apply in active systems, such as the ones found in wearable robotic garments \cite{sanchez2021textile}. Here our method provides a foundation for future work in rapidly exploring and optimizing for various use cases of active reinforcements that may have high impact for the general population. sdf




\section{Acknowledgements}
This work was supported in part
by grants from the Hasler Foundation (Switzerland) and funding from the European Research Council (ERC) under the European Union’s Horizon 2020 research and innovation programme grant agreement No 717054.


\printbibliography                

\newpage


\end{document}